\documentclass[prl,twocolumn,groupedaddress,showpacs,nofootinbib]{revtex4}
\usepackage{graphicx}
\usepackage{dcolumn}
\usepackage{amssymb}
\usepackage{mathrsfs}
\usepackage{amsmath}
\usepackage{epsfig}
\usepackage[dvips]{color}
\usepackage{hhline}

\begin{document}

\title{Cosmic matter from dark electroweak phase transition with neutrino mass generation}

\author{Pei-Hong Gu}
\email{peihong.gu@sjtu.edu.cn}

\affiliation{Department of Physics and Astronomy, Shanghai Jiao Tong
University, 800 Dongchuan Road, Shanghai 200240, China}

\begin{abstract}

We consider a dark electroweak phase transition, during which a baryon asymmetry in the dark neutrons and an equal lepton asymmetry in the dark Dirac neutrinos can be simultaneously induced by the CP-violating reflection of the dark fermions off the expanding dark Higgs bubbles. The Yukawa couplings for generating the ordinary Majorana neutrino masses can partially convert the dark lepton asymmetry to an ordinary baryon asymmetry in association with the ordinary sphaleron processes. The dark neutron can have a determined mass to serve as a dark matter particle. By further imposing a proper mirror symmetry, the Majorana neutrino mass matrix can have a form of linear seesaw while its Dirac CP phase can provide a unique source for the required CP violation.

\end{abstract}


\maketitle

\textbf{Introduction:} To realize a baryogenesis mechanism for dynamically generating the cosmic baryon asymmetry \cite{patrignani2016}, a CPT-invariant theory of particle interactions should match the Sakharov conditions: baryon number nonconservation, C and CP violation, departure from equilibrium \cite{sakharov1967}. Kuzmin, Rubakov and Shaposhnikov \cite{krs1985} pointed out the standard model (SM) could fulfil all of these conditions. In principle the SM can provide a so-called electroweak baryogenesis mechanism \cite{ckn1993}. This SM electroweak baryogenesis only depends on the SM parameters, in particular the Kobayashi-Maskawa (KM) \cite{km1973} CP violation in the quark sector and the mass of the Higgs boson \cite{shaposhnikov1987}. Unfortunately the KM CP violation is highly suppressed by the Jarlskog determinant \cite{jarlskog1985} while the light Higgs boson mass for the strongly first order electroweak phase transition \cite{shaposhnikov1987} has been ruled out experimentally.

On the other hand, the discovery of neutrino oscillation indicates that three flavors of neutrinos should be massive and mixed, while the cosmological observations imply the neutrino masses should be in a sub-eV range \cite{patrignani2016}. We can resort to the famous seesaw mechanism \cite{minkowski1977,yanagida1979,grs1979,ms1980} for naturally understanding the small neutrino masses. Due to the sphaleron processes \cite{krs1985} these seesaw models can also accommodate a leptogenesis mechanism \cite{fy1986} to produce the baryon asymmetry. Alternatively ones tried to connect the baryon asymmetry and the neutrino mass by the electroweak baryogenesis from the interactions involving the neutrinos \cite{ckn1990}. Although this attempt failed in some simple models with heavy neutral fermions \cite{ckn1990,hr1997}, it could succeed in a late neutrino mass framework \cite{hmp2005}.

The existence of dark matter (DM) poses another big challenge to the SM. It is very intriguing the dark and baryonic matter contribute comparable energy densities in our universe although their properties are so different \cite{patrignani2016}. This coincidence can be elegantly explained if the DM relic is due to a DM asymmetry and is related to the generation of the baryon asymmetry \cite{kaplan1992,kl2005,as2005,clt2005,klz2009,gsz2009,gs2010,acmz2010,dk2011,ptv2012,mpsz2012,gu2013,dm2012,pv2013,zurek2014,foot2014}. For example, we can consider a dark world parallel to our visible world and then transplant the baryogenesis mechanism in the visible world to the dark world. In the presence of a proper mirror symmetry \cite{foot2014}, the parameters in the dark world can be stringently constrained by those in the visible world. So, the dark and baryonic matter can have an equal number density, and hence the DM mass can become predictive as the nucleon mass is known. We can also construct some models to produce the baryon asymmetry and the DM asymmetry from the decays of same particles \cite{zurek2014,pv2013}. The related interactions even can be fully responsible for the neutrino mass generation \cite{gu2013}.

In this paper we shall propose a novel scenario to solve the coincidence problem between the baryonic and dark matter. Specifically we shall consider a dark electroweak symmetry breaking, during which the phase transition is strongly first order so that the CP-violating reflection of the dark fermions off the expanding dark Higgs bubbles can simultaneously create a baryon asymmetry in the dark neutrons and an equal lepton asymmetry in the dark Dirac neutrinos. The Yukawa couplings for generating the ordinary Majorana neutrino masses can participate in the production of the dark baryon and lepton asymmetries besides the conversion of the dark lepton asymmetry to an ordinary lepton asymmetry. When we further impose a proper mirror symmetry, the Majorana neutrino mass matrix can have a form of linear seesaw \cite{barr2004} while its Dirac CP phase can give a unique source for the required CP violation.

\textbf{The model:} The fields including the Higgs scalars and the chiral fermions are classified in an $SU(3)_c^{}\times SU(2)_L^{}\times U(1)_Y^{}$ ordinary sector (OS), an $SU(3)'^{}_c\times SU(2)'^{}_R\times U(1)'^{}_Y$  dark sector (DS) and a messenger sector (MS),
\begin{eqnarray}
\!\!\!\!\!\!\!\!&&\begin{array}{ll}
\textrm{OS}: 
&\phi^{}_d(+1)\,,~\phi^{}_u(-1)\,,~\phi^{}_e(+3)\,,~\phi^{}_\nu(-3)\,,\\
[1mm]&q_L^{}(0)\,,~d_R^{}(+1)\,,~u_R^{}(+1)\,,~l_L^{}(0)\,,~e_R^{}(+3)\,;\end{array}\nonumber\\
\!\!\!\!\!\!\!\!&&\begin{array}{ll}
\textrm{DS}: 
&\phi'^{}_d(-1)\,,~\phi'^{}_u(+1)\,,~\phi'^{}_e(-3)\,,~\phi'^{}_\nu(+3)\,,\\
[1mm]&q'^{}_R(0)\,,~d'^{}_L(-1)\,,~u'^{}_L(-1)\,,~l'^{}_R(0)\,,~e'^{}_L(-3)\,;\end{array}\nonumber\\
\!\!\!\!\!\!\!\!&&\begin{array}{ll}
\textrm{MS}: &\Sigma(0)\,,~\chi(+2)\,,~ \nu'^{}_L(-3)\,.\end{array}
\end{eqnarray}
Here the numbers in the brackets describe a Peccei-Quinn symmetry $U(1)_{\textrm{PQ}}^{}$ \cite{pq1977,weinberg1978,wilczek1978} for an invisible axion \cite{kim1979,svz1980,zhitnitsky1980,dfs1981}. We also assume the dark hypercharges and the baryon/lepton numbers of the dark fields are opposite to the ordinary ones, while the $[SU(2)^{}_L\times SU(2)'^{}_R]$-bidoublet Higgs scalar $\Sigma$, the gauge-singlet Higgs scalar $\chi$ and the gauge-singlet fermions $\nu'^{}_L$ carry the lepton numbers $+2$, $+1$, $-1$, respectively. The relevant Lagrangian is  
\begin{eqnarray}
\label{yukawa1}
\mathcal{L}_Y^{}\!\!&\supset&\!\!-\bar{y}_d^{}\bar{q}_L^{}\tilde{\phi}^{}_d d_R^{}-\bar{y}_u^{}\bar{q}_L^{}\phi^{}_u u_R^{} -\bar{y}_e^{}\bar{l}_L^{}\tilde{\phi}^{}_e e_R^{}- \bar{y}_\nu^{}\bar{l}_L^{}\phi^{}_\nu \nu'^c_L\nonumber\\
\!\!&&\!\!-\bar{y}'^{}_d\bar{q}'^{}_R\tilde{\phi}'^{}_d d'^{}_L-\bar{y}'^{}_u\bar{q}'^{}_R\phi'^{}_u u'^{}_L -\bar{y}'^{}_e\bar{l}'^{}_R\tilde{\phi}'^{}_e e'^{}_L- \bar{y}'^{}_\nu\bar{l}'^{}_R \phi'^{}_\nu \nu'^{}_L\nonumber\\
\!\!&&\!\!-f\bar{l}_L^{}\Sigma l'^{}_R-\left(\mu^{}_{1}\phi^\dagger_{d}\phi^{}_u +\mu^{}_{2} \phi^\dagger_{e}\phi^{}_d +\mu^{}_{3} \phi^\dagger_\nu \phi^{}_u \right)\chi\nonumber\\
\!\!&&\!\!-\left(\mu'^{}_{1}\phi'^\dagger_{d}\phi'^{}_u+\mu'^{}_{2} \phi'^\dagger_{e}\phi'^{}_d +\mu'^{}_{3} \phi'^\dagger_\nu \phi'^{}_u \right)\chi^\ast_{}-\rho^{}_1\phi^\dagger_d \Sigma \phi'^{}_u\nonumber\\
\!\!&&\!\!- \rho_2^{} \phi^\dagger_u \Sigma \phi'^{}_d  - \rho_3^{} \phi^\dagger_e \Sigma \phi'^{}_\nu - \rho_4^{}\phi^\dagger_\nu \Sigma \phi'^{}_e+\textrm{H.c.}\,,
\end{eqnarray}
where the lepton number is only allowed softly broken.

After the Higgs singlet $\chi$ develops its vacuum expectation value (VEV) for breaking the $U(1)_{\textrm{PQ}}^{}$ symmetry, we can define the dark and ordinary Higgs doublets,
\begin{eqnarray}
\phi'=\frac{\sum_{i=d,u,e,\nu}^{}\langle\phi'^{}_{i}\rangle\phi'^{}_{i}}{\sqrt{\sum_{i=d,u,e,\nu}^{}\langle\phi'^{}_{i}\rangle^2_{}}}\,, 
~~\phi=\frac{\sum_{i=d,u,e,\nu}^{}\langle\phi^{}_{i}\rangle\phi^{}_{i}}{\sqrt{\sum_{i=d,u,e,\nu}^{}\langle\phi^{}_{i}\rangle^2_{}}}\,,
\end{eqnarray}
and then get the terms for the fermion mass generation,
\begin{eqnarray}
\label{yukawa2}
\!\!\!\!\!\!\!\!  \mathcal{L}\!\!&\supset&\!\!-y_d^{}\bar{q}_L^{}\tilde{\phi} d_R^{}-y_u^{}\bar{q}_L^{}\phi u_R^{} -y_e^{}\bar{l}_L^{}\tilde{\phi}e_R^{}- y_\nu^{}\bar{l}_L^{}\phi \nu'^c_L\nonumber\\
\!\!\!\!\!\!\!\!\!\!&&\!\!-y'^{}_d\bar{q}'^{}_R\tilde{\phi}' d'^{}_L-y'^{}_u\bar{q}'^{}_R\phi' u'^{}_L -y'^{}_e\bar{l}'^{}_R\tilde{\phi}'e'^{}_L- y'^{}_\nu\bar{l}'^{}_R \phi' \nu'^{}_L\nonumber\\
\!\!\!\!\!\!\!\!\!\!&&\!\!-f\bar{l}_L^{}\Sigma l'^{}_R-\rho\phi^\dagger_{}\Sigma \phi'+\textrm{H.c.}~~\textrm{with}~~y^{}_{i}=r^{}_i\bar{y}^{}_{i}\,,\nonumber\\
\!\!\!\!\!\!\!\!\!\!&&\!\!y'^{}_{i}=r'^{}_i\bar{y}'^{}_{i}\,,
~\rho=\!\!\sum_{i,j,k}^{}r^{}_i r'^{}_j\rho_k^{}\,,~r^{}_i=\frac{\langle\phi_{i}^{}\rangle}{\langle\phi\rangle}\,,~r'^{}_i=\frac{\langle\phi'^{}_{i}\rangle}{\langle\phi'\rangle}\,\!.
\end{eqnarray} 
Note the messenger Higgs bidoublet $\Sigma$ can also acquire a VEV because of the above $\rho$-term. This Higgs bidoublet indeed can be expressed by two ordinary doublets, i.e. $\Sigma=[\sigma_1^{}~\sigma_2^{}]$ with $\langle\sigma_1^{}\rangle=0$ and $\langle\sigma_2^{}\rangle=\langle\Sigma\rangle$. We thus can have the SM Higgs doublet $H= \phi\cos\beta+ \sigma_2^{}\sin\beta$ with the rotation angle $\tan\beta = \langle\Sigma\rangle /  \langle\phi\rangle$ and the VEV $\langle H\rangle=\sqrt{\langle\phi\rangle^2_{}+\langle\Sigma\rangle^2_{}}\simeq 174\,\textrm{GeV}$. 

While the ordinary and dark charged fermion masses are produced in a usual way, the neutrino masses are suppressed by a seesaw mechanism. Specifically the dark neutrinos $\nu'^{}_R$ and the messenger fermions $\nu'^{}_L$ form three heavy dark Dirac neutrinos, meanwhile, the ordinary neutrinos $\nu^{}_L$ obtain a tiny Majorana mass term, i.e.
\begin{eqnarray}
\label{seesaw1}
\!\!\!\!\!\!\mathcal{L}\!\!&\supset&\!\!-\frac{1}{2}\!\left[\bar{\nu}_L^{}~ \bar{\nu}'^{c}_R~\bar{\nu}'^{}_L\right]\!\! \left[\!\begin{array}{ccc}0& f\langle\Sigma\rangle & y_\nu^{}\langle\phi\rangle\\
f^T_{}\langle\Sigma\rangle&0&y'^\ast_{\nu} \langle\phi'\rangle\\
y^T_{\nu} \langle\phi\rangle&y'^\dagger_{\nu} \langle\phi'\rangle&0
\end{array}\!\right]\!\!\!\left[\begin{array}{c}
\nu^c_L\\
\nu'^{}_R\\
\nu'^{c}_L
\end{array}\right]\!\!+\!\textrm{H.c.}\nonumber\\
\!\!\!\!\!\!\!\!&\simeq&\! \!-\frac{1}{2}\bar{\nu}^{}_L m_\nu^{} \nu^c_L - \bar{\nu}'^{}_R M'^{}_\nu \nu'^{}_L+\textrm{H.c.}~~\textrm{with}\nonumber\\
\!\!\!\!\! M'^{}_\nu\!\!&=&\!\!y'^{}_\nu \langle\phi'\rangle\!\gg\! m_\nu^{}\!=\!\!\left[f\frac{1}{y'^{\ast}_\nu}y^{T}_\nu\!+\!y^{}_\nu \frac{1}{y'^\dagger_\nu}f^T_{}\right]\!\!\frac{\langle H\rangle^2_{}\sin 2\beta}{2\langle\phi'\rangle}\,.
\end{eqnarray}
Apparently the Majorana neutrino masses can be suppressed by the Yukawa couplings $y^{}_\nu$ and/or the VEV ratio $\langle H \rangle \sin 2\beta /\langle\phi'\rangle$. Alternatively the Yukawa couplings $f$ can be very small though this choice is quite arbitrary.

We can introduce a mirror symmetry to simplify the parameter choice. Actually the existence of our dark sector can be well motivated by such a mirror symmetry. Here we choose the mirror symmetry to be the CP under which the gauge and Yukawa couplings are 
\begin{eqnarray}
\label{mirror}
g'^{}_{1,2,3}=g^{}_{1,2,3}\,,~~\bar{y}'^{}_{d,u,e,\nu}=\bar{y}^{\ast}_{d,u,e,\nu}\,,~~f=f^T_{}\,.
\end{eqnarray}
The Majorana neutrino mass matrix in Eq. (\ref{seesaw1})  then will be a linear seesaw \cite{barr2004}, i.e.
\begin{eqnarray}
m^{}_\nu&=&- f \frac{\langle H\rangle^2_{}\sin 2\beta }{ \langle \phi' \rangle}\frac{r^{}_\nu}{r'^{}_\nu}\nonumber\\
&=&U^\ast_{}\hat{m}_\nu^{} \textrm{diag}\{e^{-i\alpha_{1}^{}}_{},e^{-i\alpha_{2}^{}}_{},1\} U^{\dagger}_{}\,,
\end{eqnarray}
where the unitary matrix $U$ contains a Dirac CP phase.

\textbf{Dark baryon and lepton numbers:} We require the dark electroweak symmetry breaking before the ordinary electroweak symmetry breaking. Moreover the phase transition during the dark electroweak symmetry breaking is required strongly first order \cite{mr2012}. These two assumptions definitely can be achieved since we have the flexibility to choose the proper parameters in the scalar potential. We will study the details elsewhere.   

During such a dark electroweak phase transition, the bubbles of the true ground state of the dark Higgs scalar $\phi'$ will nucleate and expand until they fill the universe. Outside the bubbles where the dark electroweak symmetry is unbroken, the right and left-handed dark fermions can have distinct thermal masses and hence different momenta perpendicular to the bubble wall. Furthermore the dark $SU(2)'^{}_R$ sphaleron reactions can keep very fast outside the bubbles though they are highly suppressed inside the bubbles. As a dark Higgs bubble expands, the dark fermions from the unbroken phase will be reflected off the bubble wall back into the unbroken phase. If the CP is not conserved, we can expect a difference between the reflection probabilities for the dark quarks and antiquarks with a given chirality. We hence can obtain a net baryon number outside the bubbles and an opposite baryon number inside the bubbles. Subsequently the baryon number outside the bubbles, other than the baryon number inside the bubbles, will be converted to a lepton number and a baryon number by the dark sphaleron processes. The final baryon number thus should be a sum of the baryon number inside the bubbles and the baryon number outside the bubbles. This means the dark baryon asymmetry must equal the dark lepton asymmetry. Similarly we can consider the CP-violating reflection of the dark leptons and antileptons off the dark Higgs bubbles.

The above scenario is just an application of the SM electroweak baryogenseis mechanism in our dark sector. Note the KM-type CP violation of the dark quarks or leptons can be as large as order one because of the free Yukawa couplings $y'^{}_{d,u,e,\nu}$ and $f$. As an example we consider the lepton number from the reflection of the dark leptons and antileptons off the dark Higgs bubbles \cite{hs1995},
\begin{eqnarray}
\label{lnumber}
\!\!\!\!n^{r}_L\!\simeq \!\int\! \frac{d\omega}{2\pi}n_0^{}(\omega)\left[1-n_0^{}(\omega)\right]\frac{\Delta k \cdot v_W^{}}{T}\Delta(\omega)+\mathcal{O}(v^2_W)\,.
\end{eqnarray}
Here $\omega$ is the energy, $n_0^{}(\omega)=1/(e^{\omega/T}_{}+1) $ is the Fermi-Dirac distribution, $\Delta k$ is the difference between the right and left-handed dark lepton momenta perpendicular to the bubble wall, $\Delta(\omega)$ is the reflection asymmetry between the dark leptons and antileptons, and $v_W^{}$ is the advancing wall velocity. The reflected leptons and antileptons can diffuse before the bubble wall catches up. The typical distance from the advancing bubble wall to the reflected dark leptons and antileptons is $\sqrt{D^{}_Lt}-v_W^{}t$ with $D^{}_L$ being a diffusion constant \cite{fs1993,fs1994}. We have known in the SM, the strong interactions dominate the quark diffusion constant $D_B^{} \sim 6/T $ while the weak interactions dominate the lepton diffusion constant $D_L^{}\sim 100/T$ \cite{jpt1996,jpt1996-2}. In our model, the Yukawa couplings involving the dark leptons can be larger than the strong coupling and hence can dominate the lepton diffusion. So we can simply estimate the lepton diffusion constant $D_L^{}\sim c^{}_L/T$ with $1<  c_L^{}\lesssim 6$. Within the time $t_D^{}\sim D_L^{}/v^2_W$, the dark sphalerons can partially convert the lepton number $n^{r}_L$ to a baryon number $n^{}_B$. The remnant lepton number outside the bubbles and the lepton number inside the bubbles can give a net lepton number $n_L^{}$. By solving the diffusion equations, the baryon and lepton numbers inside the expanded bubbles should be \cite{fs1993,fs1994}
\begin{eqnarray}
\label{blnumber}
\frac{n_L^{}}{s}=\frac{n^{}_B}{s}\sim-\frac{9 \Gamma_{\textrm{sph}}^{}}{T^3_{}}\frac{D_L^{}}{v_W^2}\frac{n^{r}_{L}}{s}=-\frac{3^3_{}g'^8_2 \kappa  c_L^{}}{2^7_{}\pi^4_{}v_W^2}\frac{n^{r}_{L}}{s}\,,
\end{eqnarray}
where $\Gamma_{\textrm{sph}}^{}=6 \kappa \left[g'^2_2/(4\pi) \right]^5_{}T^4_{}$ is the dark sphaleron rate per volume with $\kappa\simeq 20$ being a coefficient \cite{bmr2000}, while $s=2\pi^2_{}g_\ast^{}T^{}_{}/45$ is the one-dimensional entropy density \cite{hs1995} with $g_\ast^{}\simeq 250.75$ being the relativistic degrees of freedom.

We now consider the mirror symmetry (\ref{mirror}) to quantitatively analyse the dark lepton and baryon numbers (\ref{blnumber}). The thermal masses of the related quasi-particles can be denoted by $\Sigma=\Omega- \gamma^0_{}(2 i  \gamma)$ with \cite{weldon1982,weldon1982-2,jpt1996,jpt1996-2,dn1995}, 
\begin{eqnarray}
\Omega_{l'^{}_R}^{2}\!&=&\!\frac{T^2_{}}{8}\left(\frac{3}{4}g^{2}_{2}+\frac{1}{4}g^2_{1}+ U\hat{f}^2_{} U^\dagger_{}+\frac{1}{2 } \hat{\bar{y}}^2_e +\frac{1}{2} \bar{y}^{\ast}_\nu \bar{y}^{T}_\nu \right)\,,\nonumber\\
\Omega_{e'^{}_L}^{2}\!&=&\!\frac{T^2_{}}{8}\left(g^2_1+ \hat{\bar{y}}^2_e \right)\,,~~\Omega_{\nu'^{}_L}^{2}=\frac{T^2_{}}{8}  \bar{y}^{T}_\nu \bar{y}^{\ast}_\nu \,,\nonumber\\
\gamma_{l'^{}_R}^{}\!&\sim &\!\frac{T}{32\pi}\left[9g^{2}_{2}+6g^2_{1}+U\hat{f}^2_{} U^\dagger_{}+\frac{1}{2}\left(\hat{\bar{y}}^2_e + \bar{y}^{\ast}_\nu \bar{y}^{T}_\nu\right) \right],\nonumber\\
\gamma_{e'^{}_L}^{}\!&\sim &\frac{T}{32\pi}\left(12 g^2_1+ \hat{\bar{y}}^2_e \right),~~\gamma_{\nu'^{}_L}^{2}\sim \frac{T}{32\pi}\left( \bar{y}^{T}_\nu \bar{y}^{\ast}_\nu\right) .
\end{eqnarray}
Because the Yukawa couplings $\bar{y}^{}_\nu$ have a totally unknown structure, we shall assume $\bar{y}_\nu^{}\ll f,\bar{y}_e^{} $ to conveniently ignore $\bar{y}^{}_\nu$ from the calculations. We shall also consider a quasi-degenerate neutrino spectrum, i.e. $\delta\!\hat{f}^2_{}=\hat{f}^2_{}-\hat{f}_3^{2}\ll \hat{f}^2_{1,2,3}$, and take $3g_2^2/4+g_1^2/4+\hat{f}_3^{2}> \hat{\bar{y}}_e^{2}/2$. Under these assumptions, we can perform
 \begin{eqnarray}
\Omega_{l'^{}_R}^{}\!\!&\simeq&\!\! \Omega_{l'^{}_R}^{(0)}+\Omega_{l'^{}_R}^{(1)}\,,~~\Omega_{l'^{}_R}^{(0)}\simeq \frac{T}{2\sqrt{2}}\sqrt{\frac{3}{4}g_2^2+\frac{1}{4}g_1^2+\hat{f}^{2}_3}\,,\nonumber\\
\Omega_{l'^{}_R}^{(1)}\!\!&\simeq&\!\!\frac{T}{4\sqrt{2}}\frac{\left(U\delta\!\hat{f}^2_{} U^\dagger_{} +\frac{1}{2}\hat{\bar{y}}^2_e\right)}{\sqrt{\frac{3}{4}g_2^2+\frac{1}{4}g_1^2+\hat{f}^{2}_3}}\,,~~\Omega_{e'^{}_L}^{}=\frac{T}{2\sqrt{2}}\sqrt{g^2_1+ \hat{\bar{y}}^2_e}\,,\nonumber\\
\Omega_0^{} \!\!&=&\!\!\frac{\Omega_{l'^{}_R}^{(0)} + \Omega_{e'^{}_L}^{}}{2}\,,~\delta p=-3 \Omega_{l'^{}_R}^{(1)}\,,~\Delta k \!=3\!\left(\Omega_{l'^{}_R}^{(0)} - \Omega_{e'^{}_L}^{}\right), \nonumber\\
\bar{\gamma}\!\!&=&\!\!\frac{\gamma_{l'^{}_R}^{}+ \gamma_{e'^{}_L}^{}}{2}\sim \frac{T}{64\pi}\left(9g_2^2+18 g_1^2+\hat{f}_3^2\right).
\end{eqnarray}
When the perturbation condition $2\Omega_{l'^{}_R,e'^{}_L}^{}> r'^{}_e\hat{\bar{y}}_e^{} \langle \phi'\rangle$ is satisfied, the reflection asymmetry $\Delta (\omega)$ can be analytically solved by \cite{hs1995}
\begin{eqnarray}
\!\!\!\!\Delta (\omega)\!\!&=&\!\!-i\frac{\textrm{Tr}\left[\left(r'^{}_e\hat{\bar{y}}_e^{} \langle \phi'\rangle\right)^2_{}, \delta p \right]^3_{}}{2^7_{}\cdot 3^{10}_{}\bar{\gamma}^{9}_{}}\left[\!1+\!\left(\frac{\omega -\Omega_0^{}}{\bar{\gamma}}\right)^2_{}\right]^{-6}_{}\nonumber\\
\!\!\!\!\!\!&=&\!\!-\frac{2^{\frac{69}{2}}_{}\pi^9_{} r'^{6}_{e} \hat{\bar{y}}_{\tau}^6 \hat{f}_3^6 \left[\!1+\!\left(\frac{\omega -\Omega_0^{}}{\bar{\gamma}}\right)^2_{}\right]^{\!-6}_{}}{\left(\frac{3}{4}g^{2}_{2}+\frac{1}{4}g^2_{1}+\hat{f}_3^2\right)^{\!\frac{3}{2}}_{}\!\!\left(9g_2^2+18 g_1^2+\hat{f}_3^2\right)^{\!9}_{}} \left(\!\frac{\langle\phi'\rangle}{T}\!\right)^{\!6}_{}\nonumber\\
\!\!\!\!\!\!\!\!&&\!\!\times \!\!\prod_{i>j \atop e,\mu,\tau}\!\!\!\left(\frac{m_{i}^{2}-m_{j}^{2}}{m_\tau^2}\right)\!\!\prod_{i>j \atop 1,2,3}^{}\!\!\!\left(\frac{m_{\nu_i^{}}^2-m_{\nu_j^{}}^2}{m_{\nu_3^{}}^2}\right) J_{CP}^{}\,,
\end{eqnarray}
with $J_{CP}^{}=\frac{1}{4} \sin 2\theta_{12}^{}\sin 2 \theta_{23}^{}\cos^2_{}\theta_{13}^{}\sin\theta_{13}^{}\sin\delta_{CP}^{}$. For the known parameters $m_{\nu_3^{}}^2-m_{\nu_1^{}}^2=2.45\times 10^{-3}_{}\,\textrm{eV}^2_{}$, $m_{\nu_2^{}}^2-m_{\nu_1^{}}^2=7.53\times 10^{-5}_{}\,\textrm{eV}^2_{}$, $\sin^2_{}\theta_{12}^{}=0.304$, $\sin^2_{}\theta_{23}^{}=0.51$, $\sin^2_{}\theta_{13}^{}=0.0219$, $m_\tau^{}=1.78\,\textrm{GeV}$, $m_\mu^{}=106\,\textrm{MeV}$, $m_e^{}=511\,\textrm{keV}$, $g_2^{}=0.653$, and $g_1^{}=0.358$ \cite{patrignani2016}, we can input $\hat{f}_3^{}=\sqrt{4\pi}$, $\hat{\bar{y}}_{\tau}^{}=\sqrt{4\pi}$, and $r'^{}_e=1/\sqrt{2}$ to estimate an upper bound of the dark baryon and lepton numbers, 
\begin{eqnarray}
\label{lnumber2}
\frac{n_L^{}}{s}&=&\frac{n^{}_B}{s}\sim 2.75\times 10^{-8}_{}\times \left(\frac{\kappa}{20}\right)\!\left(\frac{c_L^{}}{1}\right)\!\left(\frac{0.1}{v_W^{}}\right) \!\left(\frac{\langle\phi'\rangle}{T}\right)^{\!\!6}_{}\nonumber\\
&&\times \left(\frac{0.1\,\textrm{eV}}{m_{\nu_3^{}}^{}}\right)^{\!\!6}_{}\!\left(\frac{\sin\delta_{CP}^{}}{1}\right)\,.
\end{eqnarray}

\textbf{Ordinary baryon number:} The dark baryon asymmetry is expected to account for the DM relic density. We also expect the dark lepton asymmetry can be transferred to an ordinary lepton asymmetry and hence can participate in the ordinary $SU(2)^{}_L$ sphaleron processes. If the dark baryon asymmetry is stored in the dark protons, the dark lepton asymmetry should be stored in the dark electrons as a result of the neutrality of the dark electric charge. Through the Yukawa couplings, $-f\bar{l}_L^{}\sigma_1^{} e'^{}_R+\textrm{H.c.}$, the ordinary lepton doublets $l_L^{}$ and the scalar doublet $\sigma_1^{}$ can inherit this dark lepton asymmetry. So the neutral component of the $\sigma_1^{}$ scalar can keep stable to have a significant number density. However, this is not experimentally allowed because the $\sigma_1^{}$ scalar has the ordinary electroweak gauge interactions. Fortunately, we can consider another scenario that the dark baryon asymmetry is stored in the dark neutrons while the dark lepton asymmetry is stored in the dark neutrinos. The conversion between the dark and ordinary lepton asymmetries depends on the following terms, 
\begin{eqnarray}
\label{yukawa3}
\!\!\mathcal{L}\!\supset\!- y_\nu^{}\bar{l}_L^{}\phi \nu'^c_L\!-\!\!f\bar{l}_L^{}\sigma_2^{} \nu'^{}_R\!-\!\!M'^{}_\nu \bar{\nu}'^{}_R \nu'^{}_L\!\!-\!\!\rho \langle\phi'\rangle \phi^\dagger_{}\sigma^{}_{2}\!+\!\textrm{H.c.}.\end{eqnarray}
Note the above two Yukawa terms will wash out the lepton asymmetry if they are both strong enough. We hence require one of them to keep departure from equilibrium until the ordinary sphalerons stop working at the temperature $T_{\textrm{sph}}^{}\sim 100\,\textrm{GeV}$ \cite{krs1985}. For example, we can easily estimate $y^{}_\nu \lesssim \mathcal{O}(10^{-7}_{})\ll f\lesssim \mathcal{O}(1)$ or $f\lesssim \mathcal{O}(10^{-7}_{})\ll y_\nu^{} \lesssim \mathcal{O}(1)$ in the limiting case where the ordinary Higgs doublets $\phi$ and $\sigma^{}_2$ are both near the ordinary electroweak scale. 

We now derive the relation between the ordinary baryon asymmetry and the dark lepton and baryon asymmetries. At the crucial temperature $T_{\textrm{sph}}^{}$, the SM Yukawa interactions, the $SU(2)^{}_L$ and $SU(3)_c^{}$ sphalerons as well as the vanishing $U(1)^{}_Y$ hypercharge can yield some relations among the chemical potentials $\mu_{q,d,u,l,e,H}^{}$ of the SM fields  $q_L^{}, d_R^{}, u^{}_R, l^{}_L, e^{}_R, H$ \cite{ht1990,mz1992,gs1994}. Specifically, all chemical potentials can be expressed in terms of a single chemical potential. For example, we read $\mu_q^{}=-\frac{1}{3}\mu_l^{}$, $\mu_d^{}=-\frac{19}{21}\mu_l^{}$, $\mu_u^{}=\frac{5}{21}\mu_l^{}$, $\mu_e^{}=\frac{3}{7}\mu_l^{}$ and $\mu_H^{}=-\frac{4}{7}\mu_l^{}$ \cite{ht1990}. The lepton number then can be described by $L=3\left(2\mu_l^{}+\mu_e^{}\right)+N_{\nu'}^{}(\mu_{\nu'^{}_R}^{}+\mu_{\nu'^{}_L}^{})=\frac{51+22N_{\nu'}^{}}{7}\mu_l^{}~\textrm{for}~f\gg y_{\nu}^{}$ or 
$L=\frac{51-22N_{\nu'}^{}}{7}\mu_l^{}~\textrm{for}~f\ll y_{\nu}^{}$. Here $N_{\nu'}^{}=0,1,2,3$ denotes the number of the relativistic dark neutrinos, while $\mu_{\nu'^{}_{R,L}}^{}$ are their chemical potentials determined by
$\mu_{\nu'^{}_{L}}^{}=\mu_{\nu'^{}_{R}}^{}=+\frac{11}{7}\mu^{}_l~\textrm{for}~ f^{}_{}\gg y_{\nu}^{}$ or
$\mu_{\nu'^{}_{L}}^{}=\mu_{\nu'^{}_{R}}^{}=-\frac{11}{7}\mu^{}_l~\textrm{for}~ f\ll y_{\nu}^{}$.
The ordinary baryon number thus should be 
\begin{eqnarray}
\label{bnumber2}
\!\!\!\!B_f^{}\!=\!\left\{\!\!\begin{array}{lcl}\frac{28}{79+22N_{\nu'}^{}}(B-L)_{i}^{}=\!-\frac{28}{79+22N_{\nu'}^{}}B_{n'}^{}\!\!\!&\textrm{for}& \!f\gg y_{\nu}^{}\,,\\
[2mm]
\frac{28}{79-22N_{\nu'}^{}}(B-L)_{i}^{}=\!-\frac{28}{79-22N_{\nu'}^{}}B_{n'}^{}\!\!\!&\textrm{for}&\! f\ll y_{\nu}^{}\,.\end{array}\right.
\end{eqnarray} 
Here $(B-L)_i^{}=B_i^{}-L_i^{}$ is the difference between the initial values of the ordinary baryon and lepton numbers and now is just the lepton number in the dark neutrinos, equivalent to the baryon number in the dark neutrons, i.e.  $B_i^{}=0$ and $(B-L)_i^{}=- L^{}_{\nu'}=- B^{}_{n'}$.

\textbf{Predictions and implications:} By inputting the cosmological observations $\Omega_{\textrm{b}}^{}h^2_{}=0.02226$ and $\Omega_{\textrm{dm}}^{}h^2_{}=0.1186$ \cite{patrignani2016}, the dark neutron should have a determined mass to serve as the DM particle, i.e.
\begin{eqnarray}
m_{n'}^{}=m_p^{}\frac{B_f^{}}{B_{n'}^{}}\frac{\Omega_{\textrm{dm}}^{}h^2_{}}{\Omega_{\textrm{b}}^{}h^2_{}}= \frac{149}{\left|79\pm 22 N_{\nu'}^{}\right|}m_p^{}\,.
\end{eqnarray} 
After fixing $N_{\nu'}^{}=(0,1,2,3)$, we can easily read $m_{n'}^{}=(1.89,1.48,1.21,1.03)m_p^{}$ for $f\gg y_\nu^{}$, or $m_{n'}^{}=(1.89,2.61,4.26,11.5)m_p^{}$ for $f\ll y_\nu^{}$.

Our model contains a massless dark photon $\gamma'$, which should be consistent with the BBN. Now the annihilations between the dark nucleons and antinucleons into the dark pions are very fast. Actually, the cross section is much larger than the typical value $1\,\textrm{pb}$ for the usual thermally produced DM. This means the dark neutron-antineutron annihilation can be frozen out at a temperature around $T_F^{}\sim m_{n'}^{}/30$ \cite{kt1990} and then the dark pions can immediately decay into the dark photons. So, the dark photon can go out of equilibrium at the temperature $T_{\gamma'}^{}\sim m_{n'}^{}/30$. For example, we can have $T_{\gamma'}^{}\sim 30\,\textrm{MeV}$ for $m_{n'}^{}=1.03\,m_p^{}$ or $T_{\gamma'}^{}\sim 60\,\textrm{MeV}$ for $m_{n'}^{}=1.89\,m_p^{}$. The contribution of the dark photon to the additional neutrino number is $\Delta N_\nu^{}=\frac{8}{7}\left[10.75/g_{\ast}^{}\left(T_{\gamma'}^{}\right)\right]^{\frac{4}{3}}_{}$ with $g_{\ast}^{}\left(T_{\gamma'}^{}\right)$ being the  relativistic degrees of freedom at the temperature $T_{\gamma'}^{}$. For $m_{n'}^{}=1.89\,m_{p}^{}$ with $T_{\gamma'}^{}\sim 60\,\textrm{MeV}$, we know $g_{\ast}^{}\left(60\,\textrm{MeV}\right)\simeq 60$ \cite{kt1990} and hence read $\Delta N_\nu^{}=0.11$. Actually $\Delta N_\nu^{}<1$ for $T_{\gamma'}^{}> 20\,\textrm{MeV}$ \cite{kt1990}. Our dark photon may be probed by more precise measurements in the future. 

The messenger Higgs bidoublet $\Sigma$ can result in a mass mixing between the ordinary $Z$ boson and the dark $Z'$ boson. Furthermore, we can have a $U(1)^{}_Y\times U(1)'^{}_Y$ kinetic mixing at tree and loop level. This kinetic mixing will mediate the couplings of the ordinary fermions to the dark $\gamma'$ and $Z'$ bosons. Note it is unnecessary to impose the mirror symmetry (\ref{mirror}). So, the dark electroweak symmetry can be spontaneously broken near the electroweak scale. In this case, the dark $Z'$ boson may be found at the colliders. Through the $Z'$ exchange, the scattering of the dark neutrons off the ordinary nucleons can leave a distinct signal in the DM direct detection experiments \cite{tan2016,akerib2017} since the dark neutron has a determined mass.

If the mirror symmetry (\ref{mirror}) is introduced, the dark $Z'$ boson should be far above the electroweak scale. Accordingly the dark neutron cannot be directly detected. However, we can get other interesting predictions and implications. Remarkably we can take $f\gg y_\nu^{}$ to predict $-\sin\delta_{CP}^{}> 0.01$ by using Eqs. (\ref{lnumber2}) and (\ref{bnumber2}). This prediction may be verified by the neutrino oscillation experiments. Moreover the messenger Higgs bidoublet $\Sigma$ may help to test the linear seesaw at the colliders as it is allowed a TeV mass and a sizeable VEV.

\textbf{Conclusion:} We have demonstrated a strongly first order dark electroweak phase transition with the CP-violating reflection of the dark fermions off the dark Higgs bubbles can solve the coincidence problem between the baryonic and dark matter, in association with the ordinary sphaleron processes. Remarkably the Yukawa couplings for the neutrino mass generation can play an essential role in the production of the dark baryon and lepton numbers, besides the conversion of the dark lepton number to the ordinary lepton number. Our model may be tested by the running and planning colliders, cosmological observations, DM detections or neutrino oscillations.

\textbf{Acknowledgement:} This work was supported by the National Natural Science Foundation of China under Grant No. 11675100, the Recruitment Program for Young Professionals under Grant No. 15Z127060004, the Shanghai Jiao Tong University under Grant No. WF220407201, the Shanghai Laboratory for Particle Physics and Cosmology, and the Key Laboratory for Particle Physics, Astrophysics and Cosmology, Ministry of Education.

\end{document}